\documentstyle{article}
\def\txt{\textstyle}

\def\be{\begin{equation}}
\def\ee{\end{equation}}
\def\bea{\begin{eqnarray}}
\def\eea{\end{eqnarray}}
\def\beaN{\begin{eqnarray*}}
\def\eeaN{\end{eqnarray*}}
\def\ed{\end{document}}
\def\bit{\begin{itemize}}
\def\eit{\end{itemize}}

\def\sig{\sigma}

\def\lam{\lambda}

\def\Del{\Delta}
\def\del{\delta}
\def\Bg{\Bar g}
\def\hg{\hat g}
\def\k{\kappa}
\def\alf{\alpha}

\def\BD{\Bar D}

\def\di{\partial}

\def\half{{\textstyle{1 \over 2}}}
\def\~{\tilde}
\def\lag{\hat{\cal L}}
\def\m{\label}
\def\l{\left}
\def\r{\right}

\def\goto{\rightarrow}
\def\Bar{\overline}
\def\const{\rm const}

\begin{document}

\centerline{\bf THE FIELD THEORETICAL FORMULATION OF GENERAL }
\centerline{\bf RELATIVITY AND GRAVITY WITH NON-ZERO MASSES OF
GRAVITONS} \centerline{\bf }
\medskip
\centerline{\it A.N.Petrov}
\medskip
\centerline{\it Sternberg Astronomical institute, Universitetskii
pr., 13, Moscow, 119992, RUSSIA } \centerline{{\rm e-mail:
anpetrov@rol.ru}}
\medskip

\begin{abstract}
It is a review paper related to the following topics. General
relativity (GR) is presented in the field theoretical form, where
gravitational field (metric perturbations) together with other
physical fields are propagated in an auxiliary either curved, or
flat background spacetime. A such reformulation of GR is exact
(without approximations), is equivalent to GR in the standard
geometrical description, is actively used for study of theoretical
problems, and is directed to applications in cosmology and
relativistic astrophysics. On the basis of a symmetrical (with
respect to a background metric) energy-momentum tensor for all the
fields, including gravitational one, conserved currents are
constructed. Then they are expressed through divergences of
antisymmetrical tensor densities (superpotentials). This form
permits to connect a necessity to consider local properties of
perturbations, which are analyzed  in application tasks, with the
academic imagination on the quasi-local nature of the conserved
quantities in GR.  The gauge invariance is studied, and its
properties allow to consider the problem of non-localization in
exact mathematical expressions. The M/string considerations point
out to possible modification of GR, for example, by adding ``massive
terms'' including masses of spin-2 and spin-0 gravitons. A such
original modification on the basis of the field formulation of GR is
given by Babak and Grishchuk, and we present and discuss it here.
They have shown that all the local weak-field predictions of the
massive theory are in agreement with experimental data. Otherwise,
the exact non-linear equations of the new theory eliminate the black
hole event horizons and replace a permanent power-law expansion of
the homogeneous isotropic universe with an oscillator behaviour. One
variant of the massive theory allows ``an accelerated expansion'' of
the universe.

\end{abstract}

\section{Introduction}
\m{Intro}

Frequently in general relativity (GR) investigations are carried out
under assumption that perturbations of physical fields propagate in
a background given (fixed) spacetime, flat or curved, which is a
solution to the Einstein equations \cite{LL}~-~\cite{DeWitt-book}. A
majority of tasks in modern cosmology and astrophysics are just
formulated as the study of generation, propagation, evolution and
interaction on different backgrounds. Exact cosmological and black
hole solutions, for example, are used as such backgrounds. But,
linear approximation (without taking into account ``back
reaction''), and flat or strongly simplified backgrounds frequently
only are considered. Additional assumptions are used, but it is not
clear how results will be changed without these assumptions, {\em
etc}. All of these creates a necessity in a generalized (united)
description of perturbed systems in GR. It is more natural and
desirable to to present them in the framework of a field theory in a
fixed background spacetime with the following properties:
 \bit
\item Constructions are covariant and give a possibility to use
an arbitrary curved background --- a solution to GR.
\item A perturbed system is defined by a) Lagrangian (action); b)
perturbed (field) equations; c) conserved quantities, like energy,
its density, {\em etc.}
\item Gauge (inner) transformations are defined explicitly with well
described properties convenient in applications.
\item There are no restrictions in approximations, thus
perturbed equations, gauge transformations and conservation laws are
to be exact.
 \eit

Both in theoretical analysis of perturbations and in applications,
definitions of energy, momentum, angular momentum, their densities,
and conservation laws for them turn out especially important. It is
well known that conserved quantities, like energy, are not localized
in GR. This means that it is  impossible to construct unique
covariant densities of these quantities in general. A reason is in
physical foundations of GR, namely in the equivalence principle
(see, e.g., the textbook \cite{[12]}). In mathematical terms the
problem is explained by a double role of a spacetime in GR. On the
one hand, it is an arena, on which physical fields interact and
propagate; on the other hand, the spacetime itself is a dynamical
object. Because the reason is objective, sometimes an opinion arises
that the study of the conserved quantities in GR is senseless. The
author of the paper, the same as many other researches, do not
support this point of view. The fact of the non-localization cannot
suppress the notions, like energy, themselves. Without doubts,
gravitational interaction gives a contribution into a {\em total}
energy, momentum, {\em etc}. of gravitating systems \cite{[12]}. For
example, to describe a binary star system one has to include a
notion of gravitational energy as an energy of a connection.
Considering gravitational waves in a bounded domain of empty space
one can show that this domain has a total positive energy, {\em
etc.} Such examples show that conserved quantities in GR are defined
in non-contradictive manner, at least, as non-local characteristics,
and, of course, have to be examined.

{\em Total} energy-momentum, angular momentum of an asymptotically
flat spacetime have been studied in details and are continuing to be
considered. One of the main successes was the proof of the positive
energy theorem \cite{SchoenYau}. This initiated a renewed interest
to the problem. Energy-momentum and angular momentum became to be
associated with finite spacetime domains. Such quantities are called
as {\em quasi-local} ones and their examination during last
two-three decades was very successful (see a recent nice review
\cite{Szabados04} and references there in). Returning to the
cosmological problems, where conversely more frequently local
properties of perturbations are studied, we accent a necessity to
connect local and quasi-local derivations.

There are many of approaches in GR where both an evolution of
perturbations and conservation laws for them are studied. In this
paper we consider only  one of approaches, which in a more measure
satisfies the above requirements. The perturbed Einstein equations
are rewritten as follows. The linear in metric perturbation terms
are placed on the left hand side; whereas all nonlinear terms are
transported to the right hand side, and together with a matter
energy-momentum tensor are treated as a total (effective)
energy-momentum tensor $t^{(tot)}_{\mu\nu}$. This picture was
developed in a form of a theory of a tensor field with
self-interaction in a fixed background spacetime, where
$t^{(tot)}_{\mu\nu}$ is obtained by variation of an action with
respect to a background metric. Frequently it is called as a {\it
field theoretical formulation} \cite{Grishchuk92} of GR, we will
call it simply as the {\it field formulation}, and  we just review
it here. The active history of these studies was begun in 50-th of
XX century. Deser \cite{[11]} has generalized previous works  and
suggested the more clear presentation without expansions and
approximations in Minkowski spacetime. We \cite{GPP} developed the
field formulation of GR on arbitrary curved backgrounds. Advantages
of a such description was demonstrated in several applications. A
closed Friedmann world was presented as a gravitational field
configuration in Minkowski space \cite{GP86}; trajectories of test
particles at neighborhoods of event horizons of black holes were
analyzed \cite{PetrovH}; $t^{(tot)}_{\mu\nu}$ and its gauge
properties were used for a development of elements of quantum
mechanics with gravitational self-interaction \cite{PP1}, in the
frame of which some of variants of an inflation scenario were
examined \cite{PP2}; a distribution of energy in black holes
solutions was constructed \cite{PetrovNarlikar1}; it was studied
so-called the weakest falloff conditions for asymptotically flat
spacetime at spatial infinity \cite{Petrov95}. A related
bibliography of earlier works particularly can be found in
\cite{Grishchuk92}~-~\cite{GPP}. The foundation of the field
formulation of GR and some references can be also found in the
review and discussion works \cite{ZG,G}. A more full, as we know,
the modern bibliography can be found in the works
\cite{PittsSchive2001a}.

There are different possibilities to arrive at the field formulation
of GR. Deser \cite{[11]} used  a requirement: $\bullet$ {\em for a
linear massless field of spin 2 in a background spacetime it has to
be a source in the form of a total symmetrical energy-momentum
tensor of all the fields including gravitational one.} Namely this
principal was used as a basis for constructions in \cite{GPP}. It is
well known the other method, which was more clearly represented by
Grishchuk \cite{Grishchuk92} and which briefly is formulated as a
transition $\bullet$ {\em from gravistatics (Newtonian law) to
gravidynamics (Einstein's equations)}. Keeping in mind gauge
properties of the Einstein theory one obtains the field formulation
of GR as a result of $\bullet$ {\em a ``localization'' of Killing
vectors of a background spacetime, which have a sense of parameters
in an action of a gauge theory} \cite{[16]}. The way, which has an
explicit connection with the standard geometrical formulation of GR
is based on a simple $\bullet$ {\em decomposition of usual variables
of the Einstein theory onto dynamical and background quantities}
\cite{[15]}.

In section \ref{SymApproach}, a construction of the field
formulation is given on the basis of the last of the above methods.
In section \ref{ConservationLaws}, using the results of the works
\cite{PK,PK2003b} we present conservation laws in the field
formulation of GR. The conserved currents are constructed on the
basis of symmetrical energy-momentum tensor and express, thus, local
characteristics of conserved quantities. At the same time the
currents are derived as divergences of antisymmetrical tensor
densities (superpotentials), integration of which just leads to
surface integrals, which are quasi-local conserved quantities. In
section \ref{ArbitraryDecompositions}, we give a generalization of
the results of sections \ref{SymApproach} and \ref{ConservationLaws}
for various definitions of metrical perturbations and resolve
related ambiguities. The one of more desirable properties is that an
energy-momentum complex of a theory has to be free of the second
(highest) derivatives of the field variables. The energy-momentum
tensor in \cite{GPP} does not satisfy this requirement. Babak and
Grishchuk recently improved this situation \cite{BabakGrishchuk}.
Developing the approach \cite{GPP} they reformulated the field
interpretation of GR satisfying the above property; in section
\ref{FirstDerivatives} we outline their approach and details of the
method. Gauge transformation properties in the field formulation of
GR and their connection with the non-localization problem in GR are
discussed in section \ref{gauge}. The original and perspective
technique developed in work \cite{BabakGrishchuk} is naturally
generalized for constructing a gravitational theory with  gravitons
of non-zero masses in the work \cite{BabakGrishchuk1}. In its
framework Babak and Grishchuk have also found and examined static
spherically symmetric solutions in vacuum, as well as homogeneous
and isotropic cosmological solutions. All of these  are included in
section \ref{MassiveGravity}.  We discuss some questions in the last
section \ref{Discussion}. Thus, the goal of the present review is a
presentation of the field approach to GR with its development
permitting to arrive at the Babak-Grishchuk massive gravity. This
gives a possibility to understand better the properties of this
version of the massive gravity and its connection with GR. Here, we
do not consider other variants of theories with massive gravitons.

Below we give more general notations used in the paper.

\bit

\item Greek indexes mean 4-dimensional spacetime coordinates.
Small Latin indexes from the middle of alphabet
$i,\,j,\,k,\,\ldots$, as a rule, mean 3-dimensional space
coordinates; large Latin indexes $A,\,B,\,C,\,\ldots$ are used as
generalized ones for an arbitrary set of tensor densities, like
$Q^A$. Usually $x^0 = ct$, where  $c$ is speed of light; $\k = 8\pi
G/c^2$ is the Einstein constant; $(\alf\beta)$ and $[\alf\beta]$ are
symmetrization and antisymmetrization in $\alf$ and $\beta$.

\item The dynamic metric in the Einstein theory, as usual,
is denoted by $g_{\mu\nu}$  ($g = \det g_{\mu\nu}$), whereas $\Bar
g_{\mu\nu}$ ($\Bar g = \det \Bar g_{\mu\nu}$) is the background
metric; $\eta_{\mu\nu}$ is a Minkowskian metric. A hat means that a
quantity ``$\hat Q$'' is a density of the weight $+1$, it can be
$\hat Q = \sqrt{- g} Q$, or $\hat Q = \sqrt{-\Bar g} Q$, or
independently from metric's determinants, it will be clear from a
context. A bar means that a quantity ``$\Bar Q$'' is a background
one. Particular derivatives are denoted by $(\di_i)$ and
$(\di_\alf)$; $\,(D_\alf)$ is a covariant derivative with respect to
 $g_{\mu\nu}$ with the Chistoffel symbols
$\Gamma^\alf_{\beta\gamma}$; $(\BD_\alf)$ is a background covariant
derivative with respect to $\Bar g_{\mu\nu}$ with the Chirstoffel
symbols $\Bar \Gamma^\alf_{\beta\gamma}$; $\delta/\delta Q^A$ ---
the Lagrangian derivative; $\pounds_\xi Q^A = - \xi^\alf \BD_\alf
Q^A + \l. Q^A\r|^\alf_\beta \BD_\alf \xi^\beta$ --- the Lie
derivative of a generalized tensor density $Q^A$ with respect to the
vector $\xi^\alf$; for example for the contravariant vector $Q^\mu$
one has $\pounds_\xi Q^\mu = - \xi^\alf \BD_\alf Q^\mu +  \BD_\alf
\xi^\mu$.

\item $R^\alf{}_{\mu\beta\nu}$, $R_{\mu\nu}$, $G_{\mu\nu}$,
$T_{\mu\nu}$ and $R$ are the Riemannian, Ricci, Einstein and matter
energy-momentum tensors and the curvature scalar for the physical
(effective)  spacetime; $\Bar R^\alf{}_{\mu\beta\nu}$, $\Bar
R_{\mu\nu}$, $\Bar G_{\mu\nu}$, $\Bar T_{\mu\nu}$ and  $\Bar R$ are
the Riemannian, Ricci, Einstein and matter energy-momentum tensors
and the curvature scalar for the background  spacetime.
 \eit

\section{Exact perturbed Einstein equations
on an arbitrary curved background} \m{SymApproach}
\setcounter{equation}{0}

At first we define the Lagrangian for the perturbed system. Consider
the usual action of GR:
 \be S = {1 \over c} \int d^4x \hat{\cal L}^E
\equiv
 -{1 \over {2\kappa c}} \int d^4x \hat R(g_{\mu\nu})
+ {1 \over c} \int d^4x \hat{\cal L}^M (\Phi^A,~g_{\mu\nu})
\m{(a2.1)}
 \ee
where for the sake of simplicity one assumes that $\hat{\cal L}^M
(\Phi^A,~g_{\mu\nu})$ depends on the first derivatives only. Let us
write out the  Einstein equations together with the matter ones in
the form:
 \be
 {{\del {\lag}^E} \over {\del \hat g^{\mu\nu}}} =
 -{1 \over {2\k }}
 {{\del \hat R} \over {\del \hat g^{\mu\nu}}} +
 {{\del {\lag}^M} \over {\del \hat g^{\mu\nu}}} = 0\, ,
\m{(a2.2)} \ee \be
 {{\del {\lag}^E} \over {\del \Phi^A}} =
 {{\del \lag^M} \over {\del \Phi^A}} = 0\, .
\m{(a2.3)}
 \ee
Now,  define the metric and matter perturbations as
 \be \sqrt{-g}g^{\mu\nu}  \equiv \hat g^{\mu\nu}  \equiv \Bar {\hat
g^{\mu\nu}} + \hat l^{\mu\nu}\, , \qquad \Phi^A  \equiv \Bar
{\Phi^A} + \phi^A\, . \m{(a2.4)}
 \ee
The background system is described by the action:
 \be \Bar S =
{1 \over c} \int{ d^4x \Bar{{\lag}^E}} \equiv
 -{1 \over {2\k c}} \int {d^4x \Bar{\hat R}}
+ {1 \over c} \int {d^4x \Bar{{\lag}^M}}\,, \m{(a2.8)}
 \ee
and the background quantities $\Bar{\hat g^{\mu\nu}}$ and
$\Bar{\Phi^A}$ satisfy the corresponding background equations:
 \be
 -{1 \over {2\k }}  {{\del \Bar{\hat R}} \over {\del \Bar{\hat g^{\mu\nu}}}} +
 {{\del \Bar{{\lag}^M}} \over {\del \Bar{\hat g^{\mu\nu}}}} = 0,
 \qquad  {{\del \Bar{{\lag}^M}} \over {\del \Bar{\Phi^A}}} = 0\, .
\m{(a2.6)} \ee

The perturbations $\hat l^{\mu\nu}$ and $\phi^A$ now are thought as
independent dynamic variables. The perturbed system is to be
described by a corresponding Lagrangian on the background of the
system (\ref{(a2.8)}) and (\ref{(a2.6)}). Let us construct it.
Substitute the decompositions (\ref{(a2.4)}) into the Lagrangian of
the action (\ref{(a2.1)}), subtract zero's and linear in $\hat
l^{\mu\nu}$ and  $\phi^A$ terms of the functional expansion, and add
any divergence:
 \be
 {\lag}^{dyn}  =
 {\lag}^E(\Bg+l,\,\Bar \Phi+\phi) -
 \hat l^{\mu\nu}
 {{\del \Bar{{\lag}^E}} \over {\del \Bar{\hat g^{\mu\nu}}}} -
\phi^A {{\delta \Bar {{\lag}^E}}\over{\delta \Bar{\Phi^A}}} -
 \Bar{{\lag}^E} - {1 \over {2\k }}\di_\alf \hat k^\alf
=  -{1\over{2\k}}\lag^g + \lag^m\,, \m{(2.10)}
 \ee
see on functional expansions in the the book \cite{DeWitt-book}.
Lagrangians, like (\ref{(2.10)}), are called as {\it dynamical
Lagrangians} in the terminology of \cite{[15]}. Zero's term is the
background Lagrangian, whereas the linear term is proportional to
the l.h.s. of the background equations (\ref{(a2.6)}). However, one
should not to use the background equations in ${\lag}^{dyn}$  before
its variation because really ${\lag}^{dyn}$ is not less than
quadratic in the fields $\hat l^{\mu\nu}$ and $\phi^A$ in functional
expansions.

If one chooses the vector density
 \be \hat k^\alf \equiv \hat
g^{\alpha\nu}\Del^\mu_{\mu\nu} - \hat g^{\mu\nu}
\Del^\alpha_{\mu\nu}\, \m{k-KBL}
 \ee
 with the definition
  \be
\Del^\alpha_{\mu\nu} \equiv \Gamma^\alpha_{\mu\nu} - \Bar
{\Gamma}^\alpha_{\mu\nu} = \half g^{\alf\rho}\l( \BD_\mu g_{\rho\nu}
+ \BD_\nu g_{\rho\mu} - \BD_\rho g_{\mu\nu}\r)\, , \m{DeltaDef}
 \ee
where the decomposition (\ref{(a2.4)}) is used, then a pure
gravitational part in the Lagrangian (\ref{(2.10)}) is
 \bea
\lag^{g}& =&
 \hat R
(\Bar {\hat g^{\mu\nu}} + \hat l^{\mu\nu}) - \hat l^{\mu\nu}
\Bar{R}_{\mu\nu} - \Bar{\hat g^{\mu\nu}R_{\mu\nu}} +\di_\mu \hat
k^\mu \nonumber \\&=& -(\Delta^\rho_{\mu\nu} -
\Delta^\sig_{\mu\sig}\delta^\rho_\nu)\Bar D_\rho \hat l^{\mu\nu} +
(\Bar{{\hat g}^{\mu\nu}} + \hat l^{\mu\nu})
\l(\Delta^\rho_{\mu\nu}\Delta^\sig_{\rho\sig}
-\Delta^\rho_{\mu\sig}\Delta^\sig_{\rho\nu}\r)\,. \m {(a2.16)}
 \eea
It depends on only the first derivatives of the gravitational
variables $\hat l^{\mu\nu}$. In the case of a flat background the
Lagrangian (\ref{(a2.16)}) transfers to the Rosen covariant
Lagrangian \cite{[2]}. The matter part of the dynamical Lagrangian
(\ref{(2.10)}) is
 \be
 \lag^m  =
{\lag}^M\l(\Bg+ l, \,\Bar {\Phi} + \phi\r) \nonumber - \hat
l^{\mu\nu} {{\delta \Bar {{\lag}^M}}\over{\delta \Bar {\hat
g^{\mu\nu}}}}- \phi^A {{\delta \Bar {{\lag}^M}}\over{\delta
\Bar{\Phi^A}}} - \Bar{{\lag}^M}\, . \m {(a2.15)}
 \ee

The variation of an action with the Lagrangian ${\lag}^{dyn}$ with
respect to $\hat l^{\mu\nu}$ and some algebraic calculations give
the field equations in the form:
 \be \hat G^L_{\mu\nu} + \hat
\Phi^L_{\mu\nu} = \k\l({\hat t}^g_{\mu\nu} +  {\hat t}^m_{\mu\nu}\r)
\equiv \k{\hat t}^{(tot)}_{\mu\nu}\, , \m {(a2.17)}
 \ee
where the l.h.s. linear in $\hat l^{\mu\nu}$ and $\phi^A$ consists
of the pure gravitational and matter parts:
 \be
 \hat G^L_{\mu\nu}(\hat l) \equiv
{\delta \over {\delta\Bar{g^{\mu\nu}}}} \hat l^{\rho\sig}
{{\delta\Bar{\hat R}}\over{\delta \Bar{\hat g^{\rho\sig}}}} \equiv
\half \l({\Bar D_\rho}{\Bar D^\rho}\hat l_{\mu\nu} + {\Bar
g_{\mu\nu}}{\Bar D_\rho}{\Bar D_\sig}\hat l^{\rho\sig} - {\Bar
D_\rho}{\Bar D_\nu}\hat l_{\mu}^{~\rho} - {\Bar D_\rho}{\Bar
D_\mu}\hat l_{\nu}^{~\rho}\r) , \m {(a2.18)}
 \ee
 \be \hat
\Phi^L_{\mu\nu}(\hat l, \phi) \equiv -2\k {\delta \over
{\delta\Bar{g^{\mu\nu}}}} \l(\hat l^{\rho\sig} \Bar
{{\delta{\lag}^M}\over{\delta \hat g^{\rho\sig}}} + \phi^A
\Bar{{\delta {{\lag}^M}}\over{\delta {\Phi^A}}}\r)\, . \m{(a2.19)}
 \ee
The r.h.s. of Eq. (\ref{(a2.17)}) is the total symmetrical
energy-momentum tensor density
 \be {\hat t}^{(tot)}_{\mu\nu} \equiv
2{{\delta{\lag}^{dyn}}\over{\delta \Bar{g^{\mu\nu}}}} \equiv
2{{\delta}\over {\delta \Bar{g^{\mu\nu}}}}\l(-{1\over{2\k}}\lag^g +
\lag^m  \r) \equiv {\hat t}^g_{\mu\nu} +  {\hat t}^m_{\mu\nu}.
\m{(2.20)}
 \ee
In expansions, ${\hat t}^{(tot)}_{\mu\nu}$ is not less than
quadratic in $\hat l^{\mu\nu}$ and $\phi^A$ that follows from the
form of the Lagrangian ${\lag}^{dyn}$. The explicit form
of the gravitational part is
 \be {\hat t}^g_{\mu\nu} =
{1 \over \k} \l[%
\l(-\del^\rho_\mu \del^\sig_\nu +\half \bar g_{\mu\nu} \bar
g^{\rho\sig}\r)\l(\hat\Del^\alf_{\rho\sig}\Del^\beta_{\alf\beta} -
\hat\Del^\alf_{\rho\beta}\Del^\beta_{\alf\sig}\r) + \Bar D_\tau \hat
Q^\tau_{\mu\nu}\r]\, , \m{(2.20')}
 \ee
 where
 \bea 2\hat
Q^\tau_{\mu\nu} &\equiv & -\bar g_{\mu\nu} \hat
l^{\alf\beta}\Del^\tau_{\alf\beta}+ \hat l_{\mu\nu}
\Del^\tau_{\alf\beta}\bar g^{\alf\beta}- \hat l^\tau_{\mu}
\Del^\alf_{\nu\alf}- \hat l^\tau_{\nu} \Del^\alf_{\mu\alf} +\hat
l^{\beta\tau}\l( \Del^\alf_{\mu\beta}\bar g_{\alf\nu} +
 \Del^\alf_{\nu\beta}\bar g_{\alf\mu}\r)
\nonumber \\ &+ & \hat l^{\beta}_{\mu}\l( \Del^\tau_{\nu\beta}-
 \Del^\alf_{\beta\rho}\bar g^{\rho\tau}\bar g_{\alf\nu}\r)
+\hat l^{\beta}_{\nu}\l( \Del^\tau_{\mu\beta}-
 \Del^\alf_{\beta\rho}\bar g^{\rho\tau}\bar g_{\alf\mu}\r)\, .
\m{(2.20'')}
 \eea
The matter part is expressed through the usual matter
energy-momentum tensor $ T_{\mu\nu}$ of the Einstein theory as
 \bea
\hat t^m_{\mu\nu}& = &\sqrt{-\Bar g}\l[ \l(\del^\rho_\mu
\del^\sig_\nu -\half \bar g_{\mu\nu} \bar g^{\rho\sig}\r)
\l(T_{\rho\sig} - {\half}g_{\rho\sig}T_{\pi\lam}g^{\pi\lam}\r)
 -\Bar{T}_{\mu\nu}\r]\nonumber \\
&{}& -\,2{\delta \over {\delta\Bar{g^{\mu\nu}}}} \l(\hat
l^{\rho\sig} \Bar {{\delta{\lag}^M}\over{\delta \hat g^{\rho\sig}}}
+ \phi^A \Bar{{\delta {{\lag}^M}}\over{\delta {\Phi^A}}}\r)\, ,
\m{(2.20+)}
 \eea
and is also not less than quadratic in $\hat l^{\mu\nu}$ and
$\phi^A$ in expansions. At the usual description of GR the
definition of the energy-momentum tensor by $\delta \lag^E/ \delta
g^{\mu\nu}$ is senseless because it is vanishing on the Eq.
(\ref{(a2.2)}), whereas $\hat t^{(tot)}_{\mu\nu}$ defined in
(\ref{(2.20)}) is not vanishing on the field equations
(\ref{(a2.17)}). A formal reason is that in the Lagrangian
(\ref{(2.10)}) the linear terms are subtracted.

By  the definitions (\ref{(a2.19)}) and (\ref{(2.20+)}), the field
equations (\ref{(a2.17)})  can be rewritten in the form:
 \be \hat
G^L_{\mu\nu} = \k\l( {\hat t}^g_{\mu\nu} +  \delta {\hat
t}^M_{\mu\nu} \r) = \k\hat t^{(eff)}_{\mu\nu}\,
 \m{B30}
 \ee
where $\delta{\hat t}^M_{\mu\nu} \equiv {\hat t}^M_{\mu\nu} - \Bar{
{\hat t}^M_{\mu\nu}}$ is equal to $ {\hat t}^m_{\mu\nu}$ in Eq.
(\ref{(2.20+)}) without the second line. Thus $\delta{\hat
t}^M_{\mu\nu}$ can be thought as a perturbation of $\Bar{ {\hat
t}^M_{\mu\nu}} = \Bar{ {\hat T}_{\mu\nu}}$, which includes even
linear perturbations in the dynamic fields and does not follow from
the Lagrangian (\ref{(2.10)}). But now ${\hat t}^{(eff)}_{\mu\nu}$
is the source of the {\em linear gravitational field only} without
linear matter part (see Introduction).

Let us demonstrate the equivalence with the Einstein theory.
Transfer ${\hat t}^{(tot)}_{\mu\nu}$ to the l.h.s. of  Eq.
(\ref{(a2.17)}) and use the definitions (\ref{(a2.18)}),
(\ref{(a2.19)}) and (\ref{(2.20)}) with (\ref{(2.10)}):
 \bea
 &{}& \hat G^L_{\mu\nu} + \hat \Phi^L_{\mu\nu}
- \k{\hat t}^{(tot)}_{\mu\nu}\nonumber\\ & \equiv & - 2\k {{\di \Bar
{\hat g^{\rho\sig}}}\over {\di \Bar{g^{\mu\nu}}}} {{\delta}\over
{\delta {\hat l^{\rho\sig}}}} \l[ -{1 \over {2\k }} \hat
R\l(\Bar{\hat g^{\alf\beta}} + \hat l^{\alf\beta}\r) +  {\lag}^M
\l(\Bar{\Phi^A} +
\phi^A;~\Bar{\hat g^{\mu\nu}} + \hat l^{\mu\nu}\r)\r] \nonumber \\
&{}&+~ 2\k{{\delta}\over {\delta \Bar{g^{\mu\nu}}}} \l( -{1 \over
{2\k }} \Bar{\hat R} +  \Bar{{\lag}^M} \r). \m {(2.17')}
 \eea
Because the third line is proportional to the operator of the
background equations in (\ref{(a2.6)}), then Eq. (\ref{(a2.17)}), as
seen, is the Einstein equations (\ref{(a2.2)}), only in the form
with using the decompositions (\ref{(a2.4)}).

\section{Conservation laws}
\m{ConservationLaws} \setcounter{equation}{0}

At the beginning we discuss differential conservation laws on
Ricci-flat (including flat) backgrounds. One has to take into
account $\Bar{\Phi^A} \equiv 0$,  $\Bar {{\lag}^M} \equiv 0$, $\hat
\Phi^L_{\mu\nu}\equiv 0$ and use ${\del \Bar{\hat R}/}{\del
\Bar{\hat g^{\mu\nu}}} = 0$ instead of the  background equations
(\ref{(a2.6)}). Then the Lagrangian  (\ref{(2.10)}) is simplified to
\be
 {\lag}^{dyn}  =
 -{1 \over {2\k }}  \lag^g  + \lag^m =
 -{1 \over {2\k }}  \lag^g
+ {\lag}^M \l(\phi^A;~\Bar{\hat g^{\mu\nu}} + \hat l^{\mu\nu}\r),
\m{(2.22)}
 \ee
and the field equations (\ref{(a2.17)}) transform into the form of
Eqs. (\ref{(2.17')}):
 \be \hat G^L_{\mu\nu} = \k\l({\hat
t}^g_{\mu\nu} +  {\hat t}^m_{\mu\nu}\r) \equiv \k{\hat
t}^{(tot)}_{\mu\nu}\, . \m {(a2.23)}
 \ee
Thus, for  Ricci-flat backgrounds ${\hat t}^{(tot)}_{\mu\nu}$ and
${\hat t}^{(eff)}_{\mu\nu}$ coincide, and Eqs. (\ref{(2.17')}) and
(\ref{(a2.23)}) have the form announced  in Introduction. Because in
Eq. (\ref{(a2.23)}) $\Bar D_\nu{\hat G}^{L\nu}_{\mu} \equiv 0$, a
divergence of Eq. (\ref{(a2.23)}) leads to
 \be \Bar D_\nu{\hat
t}^{(tot)\nu}_{\mu} = 0\, . \m{(2.23')}
 \ee
A contraction of ${\hat t}^{(tot)}_{\mu\nu}$ with background Killing
vectors $\lam^\alf$  gives a current $\hat{\cal J}^\nu(\lam) = \hat
t^{(tot)\nu}_\mu \lam^\mu$, which is also differentially conserved:
 \be  \BD_\nu
\hat {\cal J}^\nu \l(\lam\r) \equiv \di_\nu \hat {\cal J}^\nu
\l(\lam\r)= 0 \, . \m{CCofFF}
 \ee

Integration of $\hat{\cal J}^\nu$ leads to non-local conserved
quantities,  Consider a background 4-dimensional volume $V_4$, the
boundary of which consists of timelike ``surrounding wall'' $S$ and
two spacelike sections: $\Sigma_0 := t_0 = \const$ and $\Sigma_1 :=
t_1 = \const$. Because the conservation law (\ref{CCofFF}) is
presented by the scalar density it can be integrated through the
4-volume $V_4$: $~~ \int_{V_4} \di_\mu \hat {\cal J}^\mu(\lam) d^4 x
= 0\, . $ The generalized Gauss theorem gives
 \be \int_{\Sigma_1}
\hat {\cal J}^0(\lam) d^3 x - \int_{\Sigma_0}  \hat {\cal J}^0(\lam)
d^3 x +\oint_{S}  \hat {\cal J}^\mu(\lam) dS_\mu = 0\, \m{UseGauss}
 \ee
where $dS_\mu$ is the element of integration on $S$. If in Eq.
(\ref{UseGauss}) $~ \oint_{S}  \hat {\cal J}^\mu(\lam) dS_\mu = 0\,
, $ then the quantity
 \be {\cal P}(\lam) = \int_{\Sigma} \hat
J^0(\lam)d^3 x\, \m{(1.26)}
 \ee
is conserved  on $\Sigma$ restricted by $\di\Sigma$, intersection
with $S$. In the converse case, Eq. (\ref{UseGauss}) describes
changing the quantity (\ref{(1.26)}), that is its flux through
$\di\Sigma$. It can be also assumed
 $\di\Sigma \goto \infty$.

The differential conservation laws (\ref{(2.23')}) have also a place
for backgrounds presented by Einstein spaces in Petrov's defenition
\cite{Petrov}: $ \Bar R_{\mu\nu} = \Lambda \Bg_{\mu\nu} $ where
$\Lambda$ is a constant (see \cite{GPP,AbbottDeser82,Deser87}). For
{\em arbitrary curved} backgrounds there are no conservation laws,
like (\ref{(2.23')}). Indeed, in the general case $\Bar D_\nu\l(\hat
G^{L\nu}_{\mu} + \hat \Phi_{\mu}^{L\nu}\r) \neq 0$ in
(\ref{(a2.17)}), and $\Bar D_\nu\hat G^{L\nu}_{\mu} \neq 0$ in
(\ref{B30}). The reason is that the system (\ref{(2.10)}) interacts
with a complicated background geometry determined by the background
matter fields $\Bar{\Phi^A}$. Many of cosmological solutions are
just not of the Einstein spaces.

Conservation laws for arbitrary curved backgrounds and arbitrary
displacement vectors $\xi^\alf$ were constructed in \cite{PK2003b}.
With  using the technique of canonical N{\oe}ther procedure
developed in \cite{KBL} and applied to the Lagrangian (\ref{(2.10)})
it was obtained the identity:
 \be
 {1 \over {\k}} \hat G^{L\mu}_{\nu}\xi^\nu + {1 \over
 \k} \hat l^{\mu\lam} \Bar
 R_{\lam\nu}\xi^\nu + \hat \zeta^\mu \equiv {\Bar D_\nu} \hat
 I^{\mu\nu} \equiv {\di_\nu} \hat
 I^{\mu\nu}\, .
 \m{B27'}
 \ee
The superpotential has the form: \be
 \hat I^{\mu\nu}
\equiv {1 \over \k} \hat l^{\rho[\mu}\Bar D_\rho\xi^{\nu]} + \hat
{\cal P}^{\mu\nu}{_\lambda} \xi^\lambda \equiv {1 \over \k} \l(\hat
l^{\rho[\mu}\Bar D_\rho\xi^{\nu]}+ \xi^{[\mu}\Bar D_\sig \hat
l^{\nu]\sig}-\bar D^{[\mu}\hat l^{\nu ]}_\sig \xi^\sig \r)
\m{alaAbbottDeser}
 \ee
and, thus, $\di_{\mu\nu} \hat I^{\mu\nu} \equiv 0$. It  generalizes
the Papapetrou superpotential \cite{Papapetrou48}; indeed for the
translations in Minkowski space $\xi^\lam = \delta^\lam_{(\rho)}$ in
the Lorentzian coordinates one gets
 \be
 \hat I^{\mu\nu}_{(\rho)} =
\hat {\cal P}^{\mu\nu}{}_{\rho} = {1 \over 2\k}
\di_\sig\l(\delta^{\mu}_\rho \hat l^{\nu\sig}-\delta_{\rho}^{\nu}
\hat l^{\mu\sig}-  \Bar g^{\sig\mu} \hat l^{\nu}_{\rho}+\Bar
g^{\sig\nu} \hat l^{\mu}_{\rho}\r)\, . \m{(2.27)}
 \ee
The same superpotential (\ref{alaAbbottDeser}) was constructed in
\cite{PK} by the other way, namely, by the Belinfante symmetrization
of the canonical system in \cite{KBL}. The last term on the l.h.s.
of (\ref{B27'}) is
 \be
 2\k \hat \zeta^\mu \equiv
2\l(\bar z^{\rho\sig}\Bar D_\rho \hat l^\mu_\sig - \hat
l^{\rho\sig}\Bar D_\rho \bar z^\mu_\sig\r) - \l(\bar
z_{\rho\sig}\Bar D^\mu \hat l^{\rho\sig} - \hat l^{\rho\sig}\Bar
D^\mu \bar z_{\rho\sig}\r)
 + \l(\hat l^{\mu\nu}\Bar
D_\nu \bar z - \bar z\Bar D_\nu \hat l^{\mu\nu}\r) \m{Zmu-new}
 \ee
where $2\bar z_{\rho\sig} \equiv -\pounds_\xi \Bar g_{\rho\sig}$,
and, thus, disappears on the Killing vectors of the background.

To write out  a physically sensible conservation laws from the
identity (\ref{B27'}) one has to use the field equations, which we
substitute in the form of  Eq. (\ref{B30}). Then the {\em identity}
(\ref{B27'}) transforms to the {\em equation}
 \be
 \hat
I^{\mu} \equiv {\cal T}^{\mu}_{\nu} \xi^\nu + \hat \zeta^\mu = {\Bar
D_\nu} \hat I^{\mu\nu}= {\di_\nu} \hat I^{\mu\nu}\,.
 \m{B32}
 \ee
The generalized total energy-momentum tensor density is
 \be
 \hat
 {\cal T}^\mu_{\nu} \equiv {\hat t}_{\nu}^{g\mu} +
\delta {\hat  t}_{\nu}^{M\mu}+
 {1 \over \k} \hat l^{\mu\lam} \Bar R_{\lam\nu} \equiv
 \hat t^{(eff)}_{\mu\nu} +  {1 \over \k} \hat l^{\mu\lam} \Bar R_{\lam\nu}\,
 \m{B34}
 \ee
where $\hat t^{(tot)\nu}_\mu$ is exchanging  with $\hat
t^{(eff)\nu}_\mu$, and  the interaction with the background geometry
term, $\hat l^{\mu\lam} \Bar R_{\lam\nu}$, is adding. Thus,  ${\cal
T}^\mu_{\nu}$ plays the same role as $\hat t^{(tot)\nu}_\mu$ in Eq.
(\ref{CCofFF}) if Killing vectors exist. However, the  current $\hat
I^{\mu}$ has a more general applicability, than $\hat{\cal J}^{\mu}$
in (\ref{CCofFF}): it is conserved, $\Bar D_\mu \hat I^{\mu}=\di_\mu
\hat I^{\mu}=0$, on arbitrary backgrounds and for arbitrary
$\xi^\alf$. It is important, for example, for models with
cosmological backgrounds where not only the Killing vectors are used
fruitfully (see, e.g., \cite{Traschen}).

Due to antisymmetry of the superpotential (\ref{(2.27)}) the
conserved quantity, like (\ref{(1.26)}), is expressed over a surface
integral
 \be {\cal P}(\xi) = \oint_{\di\Sigma}  \hat I^{0k}(\xi)
ds_k \m{int-surface+}
 \ee
where $ds_k$ is the element of integration on $\di\Sigma$. It is
important expression because it connects a quantity ${\cal P}(\xi)$
obtained by integration of {\em local} densities with a surface
integral playing a role of a{\em quasi-local} quantity (see
discussion in Introduction).

\section{Different definitions for perturbations}
\m{ArbitraryDecompositions}
\setcounter{equation}{0}

In GR, components of each of metrical densities
 \be
 g^{a} = \l\{ g^{\mu \nu },~g_{\mu
\nu },~\sqrt{-g}g^{\mu \nu },~\sqrt{-g}g_{\mu \nu },~(-g)g^{\mu \nu
},~\ldots \r\} \m{(1)}
 \ee
could be chosen as independent dynamic variables. In the terms of
generalized variables (\ref{(1)}) the action of GR (\ref{(a2.1)}) is
rewritten as
 \be S={1\over c}\int d^{4}x\lag^{E(a)}\equiv -{1\over
{2\k c}}\int d^{4}x\hat{R}(g^a)+{1\over c}\int d^{4}x\lag^{M}(\Phi
^{A},~g^a)\, . \m{(2)}
 \ee
Variation with respect to $g^a$ gives the gravitational equations in
a corresponding form instead of (\ref{(a2.2)}). The perturbations
could  be also defined for each of metric variables in Eq.
(\ref{(1)}) as
 \be \l\{g^a = \Bg^a + h^a \r\}~~=~~  \l\{ g_{\mu\nu}  = \Bg_{\mu\nu} +
h_{\mu\nu},~~ \hg^{\mu\nu}  = \Bar {\hg^{\mu\nu}} + \hat
l^{\mu\nu},~~ g^{\mu\nu}  = \Bg^{\mu\nu} +
r^{\mu\nu},~~\ldots\r\}\,.
 \m{B38}
 \ee
For the decomposition (\ref{B38}) following to the rules of
constructing the Lagrangian (\ref{(2.10)})  one gets
 \bea
 {\lag}^{dyn}_{(a)} & = &
 -{1 \over {2\k }}  \hat R\l(\Bar{ g^a} + h^a\r)
+  {\lag}^M \l(\Bar{\Phi^A} + \phi^A;~\Bar{ g^a} + h^a\r) \nonumber
 \\ &-&  h^a\l(
 -{1 \over {2\k }}  {{\del \Bar{\hat R}} \over {\del \Bar{ g^a}}} +
 {{\del \Bar{{\lag}^M}} \over {\del \Bar{g^a}}} \r) -
\phi^A {{\delta \Bar {{\lag}^M}}\over{\delta \Bar{\Phi^A}}} - \l(
-{1 \over {2\k }} \Bar{\hat R} +  \Bar{{\lag}^M} \r)
 -{1 \over {2\k }}\di_\nu \hat k^\nu\, .
\m{(2.10-a)}
 \eea
Its variation with respect to $h^a$ and some re-calculations give
the Einstein equations in the form (\ref{(a2.17)}):
 \be \hat
G^{L(a)}_{\mu\nu} + \hat \Phi^{L(a)}_{\mu\nu} = \k{\hat
t}^{(tot~a)}_{\mu\nu}\,. \m{B39}
 \ee
The total symmetrical energy-momentum tensor density is defined as
usual:
 \be
  {\hat t}^{(tot~a)}_{\mu\nu} \equiv
 2{{\delta\lag^{dyn(a)}}\over{\delta
 \Bar{g^{\mu\nu}}}}\, .
 \m{B43}
 \ee
In the l.h.s. of Eq. (\ref{B39}) the independent variables $h^a$ are
replaced by the other variables
 \be
 \hat l^{\mu\nu}_{(a)} \equiv
 h^a {{ \di \Bar {\hat g^{\mu\nu}}} \over {\di \Bar {g^a}}},
 \m{B40}
 \ee
which are also considered as independent ones due to the background
equations. Thus,  the same operators (\ref{(a2.18)}) and
(\ref{(a2.19)}) are applied to  $\hat l^{\mu\nu}_{(a)}$.

For some of different decompositions (\ref{B38}): $ g^a_1 = \bar
g^a_1 + h^a_1$ {\rm and} $ g^a_2 = \bar g^a_2 + h^a_2$ the variables
(\ref{B40}) differ one from another in the second order in
perturbations: $ \hat l^{\mu\nu}_{(a2)} = \hat l^{\mu\nu}_{(a1)} +
\hat \beta^{\mu\nu}_{(a)12}$. Because differences inter the linear
expressions of equations (\ref{B39}) the energy-momentum tensor
densities ${\hat t}^{(tot~a1)}_{\mu\nu}$ and ${\hat
t}^{(tot~a2)}_{\mu\nu}$ get the same differences too. Firstly, for
the case of flat backgrounds this fact was noted by Boulware and
Deser \cite{B-Deser}.

For the system (\ref{(2.10-a)}) the identity
 \be {1 \over {\k}} \hat
 G^{L(a)\mu}_{\nu}\xi^\nu + {1 \over \k} \hat l^{\mu\lam}_{(a)}
 \Bar R_{\lam\nu}\xi^\nu + \hat \zeta^\mu_{(a)} \equiv {\di_\nu}
 \hat I^{\mu\nu}_{(a)}\,
 \m{B50}
 \ee
takes a place and has exactly the form of the identity  (\ref{B27'})
with replacing $\hat l^{\mu\nu}$ with $\hat l^{\mu\nu}_{(a)}$ only.
Substituting Eq. (\ref{B39}) into the identity (\ref{B50}) we obtain
the conservation law in the form:
 \be \hat
 I^\mu_{(a)} = \l({\hat t}_{\nu}^{g(a)\mu} + {\delta\hat
 t}_{\nu}^{M(a)\mu} +
  \k^{-1} \hat l^{\mu\lam}_{(a)} \Bar R_{\lam\nu}\r)\xi^\nu + \hat
 \zeta^\mu_{(a)} =
 \hat {\cal T}_{(a)\nu}^{\mu}\xi^\nu + \hat
 \zeta^\mu_{(a)} =
  {\di_\nu}  \hat I^{\mu\nu}_{(a)}
 \m{B51}
 \ee
analogous to (\ref{B32}) with (\ref{B34}). Thus a family of
conservation laws (\ref{B51}) presents a corresponding family of
superpotentials, which can be presented in the form of the
Abbott-Deser type \cite{AbbottDeser82}:
 \be
 \hat {I}^{\mu\nu}_{(a)} = {1 \over \k} \l(\hat
l^{\rho[\mu}_{(a)}\Bar D_\rho\xi^{\nu]}+ \xi^{[\mu}\Bar D_\sig \hat
l^{\nu]\sig}_{(a)}-\bar D^{[\mu}\hat l^{\nu ]\sig}_{(a)} \xi_\sig
\r)\,.
 \m{B55}
 \ee
It is exact the formula (\ref{alaAbbottDeser}) with exchanging $\hat
l^{\mu\nu}$ by $\hat l^{\mu\nu}_{(a)}$. Indeed, the known
Abbott-Deser superpotential inter just this family with the
decomposition $g_{\mu\nu} = \Bar g_{\mu\nu} + h_{\mu\nu}$ through
the transformation (\ref{B40}) and for (anti-)de Sitter's
backgrounds. Superpotentials of this family differs one from another
due to the difference in perturbations. Otherwise, there is an
independent way. Thus in the work \cite{PK}, the generalized
Belinfane's technique has been applied for ``symmetrization'' of the
conserved quantities suggested by Katz, Bi{\v c}h\'ak and
Lynden-Bell \cite{KBL}. Their method does not depend on the choice
of the variables from the set (\ref{(1)}) and gives {\it uniquely}
 \cite{PK2003b} the superpotential (\ref{alaAbbottDeser}).
Thus, theoretically on the level of superpotentials the
Boulware-Deser ambiguity \cite{B-Deser} is resolved in favor of
$\hat l^{\mu\nu}$.

\section{Gravitational energy-momentum tensor}
\m{FirstDerivatives}
\setcounter{equation}{0}

In this section we present the results by Babak and Grishchuk
\cite{BabakGrishchuk}. Following them here and in next sections we
consider the Minkowski space with $\Bar R_{\alf\rho\beta\sig} = 0$
as the background spacetime. Our presentation is technically simpler
than that of Babak and Grishchuk \cite{BabakGrishchuk}, although, of
course, is equivalent to their one. Thus we repeat their
calculations on the basis of formulae (\ref{(2.10)}) -
(\ref{(2.17')}) simplified to the case of Eqs. (\ref{(2.22)}) -
(\ref{(2.23')}).  Also, in the work \cite{BabakGrishchuk} as
independent variables it is used $l^{\mu\nu} =  \hat
l^{\mu\nu}/\sqrt{-\Bar g}$, whereas we use $\hat l^{\mu\nu}$.
Leaving the field equations to be equivalent, this leads to a
difference in the direct definitions of the energy-momentum tensors.
However, with taking into account the field equations this
difference disappears and does not influence on results and
conclusions.

With using the definition (\ref{(a2.4)}) we present the expression
(\ref{DeltaDef}) through the gravitational variables $\hat
l^{\mu\nu}$:
 \bea \Delta^\lam_{\mu\nu}& \equiv&
\frac{1}{2\sqrt{-g}}\l[ g_{\mu\rho}\BD_\nu\hat l^{\lam\rho} +
g_{\nu\rho}\BD_\mu\hat l^{\lam\rho} - g_{\mu\alf}g_{\nu\beta}
g^{\lam\rho} \BD_\rho \hat l^{\alf\beta} \r. \nonumber\\& +&\l.
\half\l(g_{\alf\beta}\delta^\lam_\mu\BD_\nu\hat l^{\alf\beta}  +
g_{\alf\beta}\delta^\lam_\nu\BD_\mu\hat l^{\alf\beta} -
g_{\alf\beta} g_{\mu\nu} g^{\lam\rho} \BD_\rho \hat l^{\alf\beta}
\r)\r] \m{Delta-l}
 \eea
where $g_{\mu\nu}$, $\hat g^{\mu\nu}$ and $\sqrt{-g}$ are thought as
dependent on the definition (\ref{(a2.4)}). Substituting Eq.
(\ref{Delta-l}) into  Eq. (\ref{(2.20')}) with (\ref{(2.20'')}) one
finds that $\hat t^g_{\mu\nu}$ depends on the second derivatives of
$\hat l^{\mu\nu}$. After using the field equations (\ref{(a2.23)})
the second derivatives are left, but only minimally, as
 \bea &{}& \hat t_g^{\mu\nu} = \hat t_{(g-red)}^{\mu\nu} +
Q^{\alf\beta\mu\nu} (\hat t^m_{\alf\beta} - \half \Bar g_{\alf\beta}
\hat t^m_{\rho}{}^\rho) + (2\sqrt{-\Bar g})^{-1}
\BD_{\alf\beta}(\hat l^{\alf(\mu} \hat l^{\nu)\beta}- \hat
l^{\mu\nu} \hat l^{\alf\beta}); \m{Transformed-tg}\\
&{}&(\sqrt{-\Bar g})^2Q^{\alf\beta\mu\nu} \equiv \hat l^{\alf(\mu}
\Bar{\hat g}^{\nu)\beta} + \hat l^{\beta(\mu} \Bar{\hat
g}^{\nu)\alf}+ \hat l^{\alf(\mu} {\hat l}^{\nu)\beta} - \half
\Bar{\hat g}^{\mu\nu}\hat l^{\alf\beta}- \half  \hat
l^{\mu\nu}\l(\Bar{\hat g}^{\alf\beta}+\hat l^{\alf\beta}\r). \m{Q}
 \eea
The reduced part with only the first derivatives is
 \bea
\hat t_{(g-red)}^{\mu\nu}& =&  \frac{1}{4\k\sqrt{-\Bar g} }\l[ 2
\BD_\rho \hat l^{\mu\nu} \BD_\sig \hat l^{\rho\sig} - 2 \BD_\alf
\hat l^{\mu\alf} \BD_\beta \hat l^{\nu\beta}\r.\nonumber\\ & + &
g_{\alf\beta}\l( 2
 g^{\rho\sig} \BD_\rho \hat l^{\mu\alf}\BD_\sig \hat
l^{\nu\beta}  +  g^{\mu\nu}
\BD_\sig \hat l^{\alf\rho} \BD_\rho\hat l^{\beta\sig}\r) \nonumber \\
&-& 4 g_{\beta\rho} g^{\alf(\mu} \BD_\sig \hat l^{\nu)\beta}
\BD_\alf \hat l^{\rho\sig}\nonumber\\ & +&\l. {\txt \frac{1}{4}}(2
g^{\mu\delta} g^{\nu\omega} -
 g^{\mu\nu}  g^{\omega\delta} )(2 g_{\rho\alf}  g_{\sig\beta} -
 g_{\alf\beta}  g_{\rho\sig})
\BD_\delta \hat l^{\rho\sig}
 \BD_\omega \hat l^{\alf\beta}\r]\, .
\m{Reduced-tg}
 \eea
The matter part in (\ref{Transformed-tg}) has appeared due to using
the field equations (\ref{(a2.23)}).

In \cite{BabakGrishchuk} it was suggested the original way to
exclude the second derivatives from the energy-momentum tensor
without changing the field equations. The Lagrangian (\ref{(a2.16)})
was modified as follows
 \be \lag^g_{(mod)} = \lag^g + \hat
\Lambda^{\alf\beta\rho\sig}\Bar R_{\alf\rho\beta\sig}\, . \m{L-mod}
 \ee
This is  a typical way of incorporating  constraints (because $\Bar
R_{\alf\rho\beta\sig} = 0$) by means of the undetermined Lagrange
multipliers.  The multipliers $\hat \Lambda^{\alf\beta\rho\sig}$
form a tensor which depends on $\Bar g^{\mu\nu}$ and $\hat
l^{\mu\nu}$ (without their derivatives) and satisfy $\hat
\Lambda^{\alf\beta\rho\sig} = -\hat \Lambda^{\rho\beta\alf\sig}=
-\hat \Lambda^{\alf\sig\rho\beta}= \hat
\Lambda^{\beta\alf\sig\rho}$. Thus, the field equations
(\ref{(a2.23)}) do not change. Then, in a correspondence  with the
modified Lagrangian (\ref{L-mod}),  the modified energy-momentum
tensor density is
 \be \k \hat t^{g\,\mu\nu}_{(mod)} = \k\hat
t^{g\,\mu\nu} -\BD_{\alf\beta}\l(\hat \Lambda^{\mu\nu\alf\beta} +
\hat \Lambda^{\nu\mu\alf\beta}\r) \m{t-mod}
 \ee
instead of (\ref{(2.20')}). The originally undetermined multipliers
$\hat \Lambda^{\mu\nu\alf\beta}$ will now be determined. They can be
chosen in a such way that the remaning second derivatives in
(\ref{Transformed-tg}) can now be removed. The unique possibility is
$\hat \Lambda^{\mu\nu\alf\beta} = \l( \hat l^{\alf\nu}\hat
l^{\beta\mu}-\hat l^{\alf\beta}\hat l^{\mu\nu}\r)/4\sqrt{-\Bar g} $.
Thus the equations (\ref{(a2.23)}) are not changed, but they have to
be rewritten in the form
 \bea
 \hat G_{L(mod)}^{\mu\nu}& \equiv &
\hat G_L^{\mu\nu} - 2\BD_{\alf\beta}\hat \Lambda^{(\mu\nu)\alf\beta}
\nonumber\\
& \equiv &\frac{1}{\sqrt{-\Bar g}}\BD_{\alf\beta}\l[(\Bar{\hat
g^{\mu\nu}} + \hat l^{\mu\nu}) (\Bar{\hat g^{\alf\beta}} + \hat
l^{\alf\beta})- (\Bar{\hat g^{\mu\alf}} + \hat l^{\mu\alf})
(\Bar{\hat g^{\nu\beta}} + \hat l^{\nu\beta})\r]\nonumber\\
&=& \k\l({\hat t}_{g\,(mod)}^{\mu\nu} +  {\hat
t}_m^{\mu\nu}\r)\equiv \k{\hat t}_{(mod-tot)}^{\mu\nu}.\m
{(a2.23-mod)}
 \eea
Here, the r.h.s. defined as a symmetrical (metric) energy-momentum
tensor for the system (\ref{L-mod}) is the source for the
generalazed d'Alembert operator (general wave operator). Thus the
l.h.s. in (\ref{(a2.23-mod)}) is not more linear in $\hat
l^{\mu\nu}$. Because on the flat background the divergence of the
l.h.s. in (\ref{(a2.23-mod)}) is identically equal to zero, then $
\Bar D_\nu{\hat t}_{(mod-tot)}^{\mu\nu} = 0\, . $ In Eqs.
(\ref{(a2.23-mod)}), ${\hat t}_{(mod-tot)}^{\mu\nu}$ can be reduced
by the equations of motion, then they are rewritten as \be \hat
G_{L(mod)}^{\mu\nu} = \k \l[\hat t_{(g-red)}^{\mu\nu} +
Q^{\alf\beta\mu\nu} (\hat t^m_{\alf\beta} - \half \Bar g_{\alf\beta}
\hat t^m_{\rho}{}^\rho)
 +  {\hat t}^m_{\mu\nu}\r]\equiv \k \hat t_{(mod-tot-red)}^{\mu\nu}\,
\m {(a2.23-mod-red)}
 \ee
Thus, indeed on the equations of motion the energy-momentum tensor
density in (\ref{(a2.23-mod-red)}) is only with the first
derivatives of gravitational variables. Again $\BD_\nu\hat
t_{(mod-tot-red)}^{\mu\nu} = 0$. Let us show that Eq.
(\ref{(a2.23-mod-red)}) is equivalent to the usual Einstein
equations. Multiplying it by $\sqrt{-\Bar g}$, and
 using the identification (\ref{(a2.4)}), the definition
(\ref{(2.20+)}) for the flat background and the definition
(\ref{Q}), in the Lorentzian coordinates, one easily gets
 $$
\half\di_{\alf\beta}\l[(-g)(g^{\mu\nu}g^{\alf\beta}-
 g^{\mu\alf}g^{\nu\beta})\r] =
 \k(-g)\l(t_{LL}^{\mu\nu}  +T^{\mu\nu}\r)\, .
 $$
After substituting the Einstein equations $\k T^{\mu\nu} =
G^{\mu\nu}$ this equation transfers to the identity. Thus, indeed
Eq. (\ref{(a2.23-mod-red)}) is equivalent to the Einstein equations,
and one finds that $(-g)t_{LL}^{\mu\nu}$ is the Landau-Lifshitz's
pseudotensor \cite{LL}. After all, one concludes that $\hat
t_{(g-red)}^{\mu\nu}$ is the covariantized pseudotensor
$(-g)t_{LL}^{\mu\nu}/\sqrt{-\Bar g}$.

\section{Gauge invariance properties}
\m{gauge}
\setcounter{equation}{0}

Properties of the field formulation of GR under gauge
transformations follow from the usual covariant invariance
properties of GR in the geometrical description. We demonstrate it
briefly (for details see \cite{GPP,[15]}). Consider the same
solution to GR $\hat g^{\mu\nu}(x)$ and $\hat g'^{\mu\nu}(x')$
presented in two different coordinate systems: $\{x\}$ and $\{x'\}$
connected by the coordinate transformation  $x' = x'(x)$. Now let us
do a decomposition of the type (\ref{(a2.4)}) in both the cases:
$\hat g^{\mu\nu}(x)= \Bar{\hat g^{\mu\nu}}(x)+\hat l^{\mu\nu}(x)$
and $\hat g'^{\mu\nu}(x')= \Bar {\hat g^{\mu\nu}}(x')+\hat
l'^{\mu\nu}(x')$, with the {\em same} form of the background metric
$\Bar{\hat g^{\mu\nu}}$. For the solution in {\em primed}
coordinated from the points with coordinate quantities $x'$ one
displaces to the points with coordinate quantities $x$. After that
one has to compare both solutions. An analogous procedure has to be
carried out under the matter variables. assuming the coordinate
transformation in the form:
$$
x^{\prime\alf} = x^{\alf} + \xi^\alf + { 1\over {2!}}\xi^\beta
\xi^\alf_{~,\beta}+ { 1\over {3!}}\xi^\rho\l(\xi^\beta
\xi^\alf_{~,\beta}\r)_{,\rho} + \ldots\,,
$$
where $\xi^\mu$ are assumed as enough smooth, we obtain the
transformation \cite{GPP,[15]}:
 \be
 {\hat l}'^{\mu\nu} = \hat l^{\mu\nu} +
\sum^{\infty}_{k = 1}{1\over{k!}}~ \hbox{$\pounds$}_\xi^k \l(\Bar
{\hat g^{\mu\nu}} + \hat l^{\mu\nu}\r), \qquad {\phi}'^A = \phi^A +
\sum^\infty_{k = 1}{1\over{k!}}~\hbox{$\pounds$}_\xi^k
\l(\Bar\Phi^A+\phi^A\r)\, , \m{(5.9)}
 \ee
which are called as gauge (inner) transformations in the field
formulation of GR. Indeed, they do not affect both the coordinates
and the background quantities.

Firstly let us consider the field formulation of the section 
\ref{SymApproach}. It is not difficult to see that the Lagrangian
(\ref{(2.10)}) is invariant under the transformation (\ref{(5.9)})
up to a divergence on the background equations (\ref{(a2.6)}). The
expression (\ref{(2.17')}) gives a possibility to understand that
equations (\ref{(a2.17)}) are gauge invariant on themselves and on
the mentioned background equations. But the energy-momentum tensor
density (\ref{(2.20)}) (or (\ref{B30})) is not gauge invariant. Even
on the field equations one has
 \bea
 \k{{\hat t}}^{\prime(tot)}_{\mu\nu}& = & \k{\hat t}^{(tot)}_{\mu\nu}
  + \hat G^L_{\mu\nu}(l'-l)+\hat \Phi^L_{\mu\nu}(l'-l,\,
  \phi'-\phi)\, ,
  \nonumber\\
   \k{{\hat t}}^{\prime(eff)}_{\mu\nu}& = & \k{\hat t}^{(eff)}_{\mu\nu}
  + \hat G^L_{\mu\nu}(l'-l)\, .
  \m{tei-gauge}
  \eea

The transformations (\ref{(5.9)}) with $\Bar \Phi^A \equiv 0$, as it
has to be for $\Bar R_{\mu\nu\alf\beta} = 0$, are also gauge
transformations for the formulation in section
\ref{FirstDerivatives}. The Babak-Grishchuk Lagrangian is also gauge
invariant up to a divergence, and the equations
(\ref{(a2.23-mod-red)}) are gauge invariant on themselves.
Concerning the energy-momentum tensor, on the field equations one
has
 \be
  \k{{\hat t}}'^{(mod-tot-red)}_{\mu\nu} =
  \k{\hat t}^{(mod-tot-red)}_{\mu\nu}
  + \hat G^{L(mod)}_{\mu\nu}(l'-l).
  \m{tei-gaugeBG}
  \ee

The non-localization problem of energy and other quantities in GR
became evident from the moment of constructing GR beginning from the
original Einstein's works. During prolonged time it was illustrated
by a non-covariance of pseudotensors. The use of an auxiliary
background spacetime permitted to consider covariant conserved
quantities, but the non-localization  became to be explained by an
ambiguity in a choice of a background. However, on this level there
was no suggested an unique mechanism for description of this
ambiguity. The use of the gauge transformation properties in the
field formulation of GR closes this gap in GR. At the beginning of
this section we just express the connection between different
choices of backgrounds explicitly. A gauge non-invariance in the
energy-momentum tensors (\ref{tei-gauge}) and (\ref{tei-gaugeBG})
just expresses the non-localization of energy, momentum, {\em etc.}
in GR in exact  (without approximations) and explicit
mathematical expressions. It is a one of advantages in using the
field formulation of GR.

\section{Gravity with non-zero masses of gravitons}
\m{MassiveGravity}
\setcounter{equation}{0}

Babak and Grishchuk using their technique \cite{BabakGrishchuk} have
constructed a variant of theory of gravity with non-zero masses of
gravitons with interesting properties \cite{BabakGrishchuk1}.
Following to \cite{BabakGrishchuk1} independent variables
$l^{\mu\nu} = \hat l^{\mu\nu}/\sqrt{-\Bar g}$ are used. It is
natural to assume that the Lagrangian may also include an additional
term similar to the one in Eq. (\ref{L-mod}), but where the quantity
$\~R_{\alf\rho\beta\sig}$ is the curvature tensor of an abstract
spacetime with a constant non-zero curvature: $ \widetilde
R_{\alf\rho\beta\sig}  = K\l(\widetilde g_{\alf\beta} \widetilde
g_{\rho\sig}- \widetilde g_{\alf\sig}\widetilde g_{\rho\beta}\r) $
where $K$ is with the dimensionality of $[length]^{-2}$. If one adds
$\hat \Lambda^{\alf\beta\rho\sig}\widetilde R_{\alf\rho\beta\sig}$
with $\hat \Lambda^{\mu\nu\alf\beta} = (4\sqrt{-\widetilde
g})^{-1}\l( \hat l^{\alf\nu}\hat l^{\beta\mu}-\hat l^{\alf\beta}\hat
l^{\mu\nu}\r) $, changing $\widetilde g^{\mu\nu} \goto \Bar
g^{\mu\nu}$, then the additional term in the Lagrangian
(\ref{L-mod}) is $ \half  \sqrt{-\Bar
g}K\l(l^{\alf\beta}l_{\alf\beta}-l^{\alf}_{\alf}
l^{\beta}_{\beta}\r)$.
 Clearly, the new theory is not GR, but one recognizes in this term
the Fierz-Pauli mass-term \cite{FierzPauli}. Thus, noting that the
structure  (\ref{L-mod}) generates mass terms and finding that only
two independent quadratic combinations of $l^{\mu\nu}$ exist, Babak
and Gishchuk arrive at a 2-parametric family of theories with the
additional mass terms in the gravitational Lagrangian (\ref{L-mod}):
 \be \lag^g_{(mass)} = \lag^g_{(mod)}+ \sqrt{-\Bar
g}\l[k_1l^{\alf\beta}l_{\alf\beta}+k_2(l^{\alf}_{\alf})^{_2}\r],
\m{L-mass}
 \ee
where $k_1$ and $k_2$ have a dimensionality  of $[length]^{-2}$.

Of course, the additional term in (\ref{L-mass}) gives a
contribution both into the r.h.s. and into the l.h.s. of Eq.
(\ref{(a2.23-mod-red)}), and the equations of the new gravity theory
symbolically could be rewritten as
 \be \hat
G_{L(mass)}^{\mu\nu} =
 \k \hat t_{(tot-mass)}^{\mu\nu}\, .
\m {Eqs-mass}
 \ee
These equations are, of course, covariant, however, unlike Eq.
(\ref{(a2.23)}) and Eq. (\ref{(a2.23-mod-red)}), the new field
equations (\ref{Eqs-mass}) are not gauge invariant. There are no
transformations, like (\ref{tei-gauge}). Thus, there is no a problem
with a localization of $\hat t_{(tot-mass)}^{\mu\nu}$
--- it is localized!

To have a direct comparison with GR effects it is more convenient to
present Eq. (\ref{Eqs-mass}) in the quite equivalent
quasi-geometrical form:
 \be G_{\mu\nu} + M_{\mu\nu} = \k T_{\mu\nu}
\m{QuasiGeom}
 \ee
where the massive term  is
 $$ M_{\mu\nu} \equiv
\l(2\delta^\alf_\mu\delta^\beta_\nu -
 g^{\alf\beta}g_{\mu\nu}\r) (k_1l_{\alf\beta} + k_2\Bar g_{\alf\beta}
 l^\rho_\rho).
 $$
Notice that the Bianchi identity $D_\nu G^\nu_\mu \equiv 0$ takes a
place in effective spacetime. Besides, with taking into account the
matter equations (\ref{(a2.3)}) one has $D_\nu T^\nu_\mu = 0$, as
usual. Thus, after differentiation of Eq. (\ref{QuasiGeom}) one
obtains $D^\nu M_{\mu\nu} = 0$. Although these equations are merely
the consequences of the full system (\ref{QuasiGeom}), and therefore
contain no new information, it proves convenient to use them instead
of some members of the original set (\ref{QuasiGeom}).

To give a physical interpretation of $k_1$ and $k_2$, following to
the analysis by Ogievetsky and Polubarinov
\cite{OgievetskyPolubarinov}, and by van Dam and Veltman
\cite{VanDamVeltman}, one considers the linearization of the Eqs.
(\ref{QuasiGeom}):
 \be \half \l({\Bar D_\rho}{\Bar D^\rho}l_{\mu\nu}
+ {\Bar g_{\mu\nu}}{\Bar D_\rho}{\Bar D_\sig} l^{\rho\sig} - {\Bar
D_\rho}{\Bar D_\nu}l_{\mu}^{~\rho} - {\Bar D_\rho}{\Bar
D_\mu}l_{\nu}^{~\rho}\r) + 2k_1 l_{\mu\nu} - (k_1+2k_2) \Bar
g_{\mu\nu}l^{\alf}_{\alf}= 0. \m{LinearMass}
 \ee
The divergence of this equation is
 \be \BD_\nu\l[2k_1 l^{\mu\nu} -
(k_1+2k_2) \Bar g^{\mu\nu}l^{\alf}_{\alf}\r]= 0, \m{DivLinearMass}
 \ee
which is the linearized version of the equation $D^\nu M_{\mu\nu} =
0$.

Consider the first case with $k_1 \neq k_2$. The full system
(\ref{LinearMass}) is equivalent to
 \bea
\Box H^{\mu\nu} +\alf^2 H^{\mu\nu}& =& 0,\m{Box-H}\\
\Box l^\alf_\alf +\beta^2 l^\alf_\alf & =& 0, \m{Box-l}
 \eea
together with Eq. (\ref{DivLinearMass}). Here, $\Box \equiv \Bar
g^{\alf\beta}\BD_{\alf\beta}$,
 \be H^{\mu\nu} \equiv h^{\mu\nu}
-\frac{k_1+k_2}{3k_1} \Bar g^{\mu\nu} l^\alf_\alf -
\frac{k_1+k_2}{6k_1^2} \BD^{\mu\nu} l^\alf_\alf +
\frac{k_1+k_2}{12k_1^2} \Bar g^{\mu\nu} \Box l^\alf_\alf \m{H}
 \ee
with $ \Bar g_{\mu\nu}H^{\mu\nu} = 0$ and  $\BD_\nu H^{\mu\nu} = 0$.
Thus, parameters in the wave-like equations (\ref{Box-H}) and
(\ref{Box-l}) are
 \be \alf^2 = 4k_1\, , \qquad \qquad \beta^2 =
-\frac{2k_1(k_1 +4k_2)}{k_1 +k_2}\, . \m{AlfBeta}
 \ee
They can be thought as inverse Compton wavelengths of the {\it
spin-2} graviton with the mass $m_2 = \alf\hbar/c$ associated with
the field $H^{\mu\nu}$ and of {\it spin-0} graviton with mass $m_0 =
\beta\hbar/c$ associated with the field $l^\alf_\alf$.

With studying the weak gravitational waves in the massive gravity
one finds certain modifications of GR. Thus the spin-0 gravitational
waves, presented by the trace $l^\alf_\alf =
l^{\alf\beta}\eta_{\alf\beta}$, and the polarization state of the
spin-2 graviton presented by the spatial trace $H^{ik}\eta_{ik}$
both, unlike GR, become essential. They provide additional
contributions to the energy-momentum flux carried by the
gravitational wave, and the extra components of motion of the test
particles. However, gravitational wave solutions, their
energy-momentum characteristics, and observational predictions of GR
are fully recovered in the massless limit $\alf \goto 0$, $\beta
\goto 0$.

For the case with the mass term of Fierz-Pauli type, $k_1 +k_2 = 0$,
that corresponds $\beta^2 \goto \infty$ (see (\ref{AlfBeta})), the
full set of equations (\ref{LinearMass}) is equivalent to
$$
l^\alf_\alf = 0, \qquad \Box l^{\mu\nu} +4k_1 l^{\mu\nu} =0, \qquad
\BD_\nu l^{\mu\nu} = 0.
$$
This case is interpreted as unacceptable \cite{BabakGrishchuk1}.
Even in the limit $\alf \goto 0$, there remains a nonvanishing
``comoving mode'' motion of test particles in the plane of the wave
front. The extra component of motion is accounted  for the
corresponding additional flux of energy from the source, typically,
of the same order of magnitude as the GR flux. This, at least, is in
a conflict with already available inderect gravitational-wave
observations of binary pulsars \cite{Taylor}. Such theories probably
have to be rejected.

In \cite{BabakGrishchuk1}, the full non-linear equations
(\ref{QuasiGeom}) were analyzed from the point view of the black
hole and the cosmological solutions. Thus, searching for static
spherically-symmetric solutions in vacuum it is necessary to
consider three independent equations from (\ref{QuasiGeom}), unlike
GR where there are two ones only. The consideration is simplified if
one assumes $\alf = \beta$, however all the qualitative conclusions
remain valid for $\alf \neq \beta$. Combining analytical and
numerical technique Babak and Grishchuk have demonstrated that the
solution of the massive theory is practically indistinguishable from
that of GR for all  $2M \ll R \ll 1/\alf$,  where $R$ and $M$ are
the radial and mass parameters of the Schwarzschild solution. For
$R$ larger than $1/\alf$ the solution takes the form of the
Yukawa-type potentials; therefore they call this massive theory as
finite-range gravity. The solution of new theory is also deviate
strongly from that of GR in the vicinity of $R= 2M$ that is the
location of the globally defined event horizon of the Schwarzschild
black hole in GR.  In the massive gravity the event horizon does not
form at all, and the solution smoothly continues to the region
$R<2M$ and terminates at $R=0$ where the curvature singularity
develops. Since the $\alf M$ can be extremely small, the redshift of
the photon emitted at $R= 2M$ can be extremely large, but it remains
finite in contrast with GR solutions.  Infinite redshift is reached
only at the singularity $R=0$. In the astrophysical sense, all
conclusions that rely specifically on the existence of the black
hole even horizon, are likely to be abandoned.  It is very
remarkable and surprising that the phenomena of black hole should be
so unstable with respect to the inclusion of the tiny mass-terms,
whose Compton wavelength  can exceed, say, the present-day Hubble
radius.

It was also considered homogeneous isotropic solutions in the
framework of the massive gravity. Matter sources were taken in the
simplest form of a perfect  fluid with a fixed equation of state.
There  are two independent field equations from the set
(\ref{QuasiGeom}), unlike GR where there is only one in the same
case.  First, if the mass of the {\it spin-0} graviton is zero,
$\beta^2 = 0$, the cosmological solutions are exactly the same as
those of GR, independently of the mass of the {\it spin-2} graviton,
i.e., independently of the value of $\alf^2$. This result is
expected due to the highest spatial symmetry: the {\it spin-2}
degrees of freedom have no chance to reveal themselves. Then, for
$\beta^2 \neq 0$ it was considered technically more simple case
$4\beta^2 = \alf^2$ which was studied in full details. Qualitative
results are valid for $4\beta^2 \neq \alf^2$. Again, combining
analytical approximations and numerical calculations it was
demonstrated that the massive solution has a long interval of
evolution where it is practically indistinguishable from the
Friedmann solution of GR. The deviation from GR are dramatic at very
early times and very late times. The unlimited expansion is being
replaced by a regular maximum of the scale factor, whereas the
singularity is being replaced by a regular minimum of the one. The
smaller $\beta$, the higher maximum and the deeper minimum, i.e.,
the arbitrary small term in the Lagrangian (\ref{L-mass}) gives rise
to the oscillatory behaviour of the cosmological scale factor.

Following the logic of interpretation that $\alf^2$ and $\beta^2$
define the masses, they are thought as positive. However, the
general structure of the Lagrangian (\ref{L-mass}) does not imply
this. Then, if one allows $\alf^2$ and $\beta^2$ to be negative, the
late time evolution of the scale factor presents an ``accelerated
expansion'' that is similar to the one governed by a positive
cosmological $\Lambda$-term. The development of this point could be
useful in the light of the modern cosmological observational data
\cite{Chernin}.

\section{Concluding remarks}
\m{Discussion}
\setcounter{equation}{0}

Due to the non-localization problem/property a definition and a
study of conserved quantities in GR are not trivial tasks. Then
rather one has not to follow unconditionally to some one {\em
unique} method. However, to restrict such methods, one has to
examine their possibilities to satisfy known natural tests. As a
rule, in applications of expressions of such approaches it is
required: a) the energy density for the weak gravitational waves has
to be positive \cite{LL}; b) ratio of mass to angular momentum in
the Kerr solution has to be standard \cite{KBL}; c) one has to
obtain the standard conserved quantities both at spatial and at null
infinity for asymptotically flat solutions \cite{[12]}. The usual
field formulation (UFF) of GR presented in sections
\ref{SymApproach} and \ref{ConservationLaws}, and the
Babak-Grishchuk modification presented in section
\ref{FirstDerivatives}, they both satisfy all these evident
requirements, and one cannot do a choice on this basis.

In Introduction it was presented reasons why it is important to
study perturbations on arbitrary curved backgrounds, and to present
conserved currents as divergences of superpotentials. In particular,
just these requirements initiated a development of UFF. Conversely,
the opinion of Babak and Grishchuk is that it is enough to use {\em
only} the background Minkowski space. This has its own convincing
foundation. All the modern ``direct'' experiments, as well as a
aforementioned theoretical restrictions a) - c), use as a background
a flat spacetime. Besides, the field formulation even can describe
arbitrary curved and topologically non-trivial solutions of GR as a
field configuration in Minkowski space (see \cite{GP86,PetrovH}).
Paying a necessary opinion to the position of these authors,  we
note that there are no principal arguments against a following
generalization of the modified Babak-Grishchuk formulation of GR
onto arbitrary curved backgrounds and constructing conservation laws
with the use of superpotentials.

Let us return to the requirement to have only the first derivatives
in the energy-momentum tensor, one of the main reasons of which is a
correct formulation of the initial problem. However, at least, from
this point of view second derivatives, which appear  in the
framework of UFF  do not initiate a criticism. One needs to consider
the energy-momentum tensor density $\hat t^{g}_{\mu\nu}$ and  the
conserved current ${\hat I}^\mu$ (see (\ref{(a2.16)}) and
(\ref{B32}) with (\ref{B34})). In \cite{PK2003b} we have shown that
the symmetrized quantities constructed in \cite{PK} coincide with
the ones presented here and included into the conservation law
(\ref{B32}). Due to this, and using the dynamic and background
equations one concludes that zero's component ${\hat I}^0$ of the
conserved current in (\ref{B32}), based on $\hat t^{g}_{\mu\nu}$,
contains only the first {\it time} derivatives of $\hat l^{\mu\nu}$.
Therefore ${\hat I}^0$ has the normal behaviour with respect to
initial conditions with the definition of the integral quantities
analogous to (\ref{(1.26)}). Therefore this requirement, at least,
may be unnecessarily restrictive.

Babak and Grishchuk give and discuss a wide bibliography of works,
where a possibility to consider gravitons with non-zero masses is
studied \cite{BabakGrishchuk1}. Here, we note only the Visser paper
\cite{Visser}. It is clear that an inclusion of non-zero masses of
gravitons leads to a non-Einstein gravity. Visser, analyzing
foundations of GR, find that a more economical way turns out an
inclusion of a background metric, and he realizes this possibility.
Thus a philosophy of the works \cite{BabakGrishchuk1} and
\cite{Visser} coincide. An advantage of the approach in 
\cite{BabakGrishchuk1} is that, using the field formulation of GR,
it permits to study the problem from more general positions. Thus,
the linear Visser's equations are a particular case of Babak and
Grishchuk's linear equations, when one sets  $\alf = \beta$.
Beginning from the second order, equations in \cite{BabakGrishchuk1}
and \cite{Visser} are different even for $\alf = \beta$. The reason
is that Visser defines perturbations as $h_{\mu\nu} = g_{\mu\nu} -
\Bar g_{\mu\nu}$, whereas Babak and Grishchuk use $l^{\mu\nu}$ in
correspondence with our above definitions. It is interesting that
the same reason initiates differences in definitions of
energy-momentum tensors (see section \ref{ArbitraryDecompositions}).
In spite of these differences both of approaches,
\cite{BabakGrishchuk1} and \cite{Visser}, qualitatively give the
same predictions.

\ed
\begin{thebibliography}{999}


\bibitem{LL} L.~D.~Landau \& E.M. Lifshitz,
{\it The classical theory of fields}, (Addison-Wesley and Pergamon,
1971).

\bibitem{[12]}
Ch.~W.~Misner, K.~S.~Thorne \& J.~A.~Wheller, {\it Gravitation,}
(San Francisco, Freeman, 1973).

\bibitem{DeWitt-book}  B.~S.~DeWitt,
{\it Dynamical theory of groups and fields}, (New York, Gordon and
Breach, 1965).

\bibitem{SchoenYau} R.~Schoen \& S.-T.~Yau,
{\it Commun. Math. Phys.} {\bf 65}, 45 (1979); {\bf 79}, 231 (1981).

\bibitem{Szabados04} L.~B.~Szabados,
{\it  Living Reviews in Rela\-tivity}, {\bf 7}, 4 (2004);
www.livingreviews.org/lrr-2004-4.

\bibitem{Grishchuk92} L.~P.~Grishchuck, in: {\it Carrent Topics in
Astrofundamental Physics}, Eds.: Sanches N. and Zichichi A. (World
Scientific, 1992) p.p. 435 - 462.

\bibitem{[11]} S.~Deser, {\it Gen. Relat. Grav.} {\bf 1}, 9 (1970);
{\sf arXiv:gr-qc/0411023}.

\bibitem{GPP} L.~P.~Grishchuck, A.~N.~Petrov \& A.~D.~Popova, {\it
Commun. Math. Phys.} {\bf 94},  379 (1984).

\bibitem{GP86} L.~P.~Grishchuk \& A.~N.~Petrov, {it Pis'ma in Astron. Zh.}
 {\bf 12}, 429 (1986) [{\it Sov. Astronom. Lett.} {\bf 12}, 179
 (1986)].

\bibitem{PetrovH} A.~N.~Petrov,
{\it Astron. Astrophys. Trans.} {\bf 1}, 195 (1992).

\bibitem{PP1} A.~D.~Popova \&  A.~N.~Petrov,
    {\it Int. J. Mod. Phys. A}
{\bf 8}, 2683 (1993); {\bf 8}, 2709 (1993).

\bibitem{PP2} A.~N.~Petrov \& A.~D.~Popova,
   {\it Int. J. Mod. Phys. D}
{\bf 3}, 461 (1994); {\it Gen. Relat. Grav.} {\bf 26}, 1153 (1994).

 \bibitem{PetrovNarlikar1} A.~N.~Petrov \& J.~V.~Narlikar,
{\it Found. Phys.} {\bf{26}}, 1201 (1996), Erratum, {\it Found.
Phys}, {\bf{28}}, 1023 (1998); A.~N.~Petrov, to appear in {\it
Found. Phys. Lett.}; {\sf arXiv:gr-qc/0503082}.

\bibitem{Petrov95} A.~N.~Petrov,
{\it Int. J. Mod. Phys. D} {\bf 4}, 451 (1995); {\bf 6}, 239 (1997).

\bibitem{ZG} Ya.~B.~Zeldovich \& L.~P.~Grishchuk,
{\it Usp. Fiz. Nauk} {\bf 149}, 695 (1986) [{\it Sov. Phys. Uspekhi}
{\bf 29}, 780 (1986)]; {\bf 155}, 517 (1988) [{\bf 31}, 666 (1989)].

\bibitem{G}  L.~P.~Grishchuk, {\it Usp. Fiz. Nauk} {\bf 160}, 147 (1990) [{\it
Sov. Phys. Uspekhi} {\bf 33}, 669 (1990)].

\bibitem{PittsSchive2001a} J.~B.~Pitts \& W.~C.~Schieve,  {\it Gen. Relat.
Grav.} {\bf 33}, 1319 (2001); {\sf arXiv:gr-qc/0101058}; {\it Found.
Phys.} {\bf 34}, 211 (2004); {\sf arXiv:gr-qc/0406102}.

\bibitem{[16]} A.~N.~Petrov,
{\it Class. Quantum Grav.} {\bf 10}, 2663 (1993).

\bibitem{[15]} A.~D.~Popova \&  A.~N.~Petrov,
    {\it Int. J. Mod. Phys. A}
{\bf 3}, 2651 (1988).

\bibitem{PK} A.~N.~Petrov \& J.~Katz,
{\it Proc. R. Soc. London A} {\bf 458}, 319 (2002); {\sf
arXiv:gr-qc/9911025}.

\bibitem{PK2003b} A.~N.~Petrov,
 {\it Vestnik Mosk. Univ.
Fiz. Astron.} {\bf No. 1}, 18 (2004); [a translation to appear in
{\it Moscow Univers. Phys. Bull.}]); {\sf arXiv:gr-qc/0402090}.

\bibitem{BabakGrishchuk} S.~V.~Babak \& L.~P.~Grishchuk,
{\it  Phys. Rev. D} {\bf 61}, 24038 (2000); {\sf
arXiv:gr-qc/9907027}.

\bibitem{BabakGrishchuk1} S.~V.~Babak \& L.~P.~Grishchuk,
{\it Int. J. Mod. Phys. D} {\bf 12}, 1905 (2003); {\sf
arXiv:gr-qc/0209006}.

\bibitem{[2]} N.~Rosen,
{\it Phys. Rev.} {\bf 57}, 147, 150 (1940).

\bibitem{Petrov}  A.~Z.~Petrov,  {\it Einstein spaces}
(London, Pergamon Press,  1969).

\bibitem{AbbottDeser82}
L.~F.~Abbott \& S.~Deser, {\it Nuclear Phys. B} {\bf 195}, 76
(1982).

\bibitem{Deser87} S.~Deser,
{\it Class. Quantum Grav.} {\bf 4}, 99 (1987).

 \bibitem {KBL} J.~Katz, J.~Bi\v c\'ak \& D.~Lynden-Bell,
{\it  Phys. Rev. D} {\bf 55}, 5957 (1997); {\sf
arXiv:gr-qc/0504041}.

\bibitem {Papapetrou48} A.~Papapetrou,
{\it Proc. R. Irish Ac.} {\bf 52}, 11  (1948).

\bibitem{Traschen} J.~Traschen, {\it Phys. Rev. D} {\bf 31}, 283 (1985);
J.~Traschen \& D.~M.~Eardley, {\it Phys. Rev. D} {\bf 34}, 1665
(1986); J.-P.~Uzan, N.~Deruelle \& N.~Turok, {\it Phys. Rev. D} {\bf
57}, 7192 (1998).

\bibitem{B-Deser} D.~C.~Boulware \&  S.~Deser,
{\it Ann. Phys.} {\bf 89}, 193 (1975).

\bibitem{FierzPauli} M. Fierz \& W. Pauli, {\it Proc. R. Soc. A},
{\bf 173}, 211 (1939).

\bibitem{OgievetskyPolubarinov} V.~I.~Ogievetsky \& I.~V.~Polubarinov,
{\it Ann. Phys.} {\bf 35}, 167 (1965).

\bibitem{VanDamVeltman} H.~van~Dam \& M.~Veltman,
{\it Nucl. Phys. B} {\bf 22}, 397 (1970).

\bibitem{Taylor} J.~H.~Tailor, {\it Rev. Mod. Phys.}
{\bf 66}, 711 (1994).

\bibitem{Chernin} A.~D.~Chernin, {\it Uspekhi Fiz. Nauk}
{\bf 171}, 1153 (2001).

\bibitem{Visser} M.~Visser, {\it Gen. Relat. Grav.} {\bf 30},
1717 (1998); {\sf arXiv:gr-qc/9705051}.

\end{thebibliography}
